\documentclass[twocolumn,amsmath,amssymb,prl,10pt,nofootinbib,superscriptaddress]{revtex4}
\usepackage{ graphicx, float,amsmath, amsmath, amssymb}
\usepackage[export]{adjustbox}

\def\be{\begin{equation}}
\def\ee{\end{equation}}
\def\bea{\begin{eqnarray}}
\def\eea{\end{eqnarray}}
\def\bse{\begin{subequations}}
\def\ese{\end{subequations}}

\newcommand{\corr}[1]{{\color{red}#1}}
\usepackage[breaklinks, colorlinks, citecolor=blue]{hyperref}
\usepackage[labelsep=period]{caption}
\usepackage[normalem]{ulem}
\usepackage{mathtools}
\usepackage{siunitx}
\usepackage{soul}
\usepackage{minitoc}

\usepackage{bm}

\DeclarePairedDelimiterXPP\BigOSI[2]%
  {\mathcal{O}}{(}{)}{}%
  {\SI{#1}{#2}}

\begin{document}
\title{Closer look at white hole remnants}

\author{Aurélien Barrau}%
\affiliation{%
Laboratoire de Physique Subatomique et de Cosmologie, Universit\'e Grenoble-Alpes, CNRS/IN2P3\\
53, avenue des Martyrs, 38026 Grenoble cedex, France
}

\author{L\'eonard Ferdinand}%
\affiliation{%
Laboratoire de Physique Subatomique et de Cosmologie, Universit\'e Grenoble-Alpes, CNRS/IN2P3\\
53, avenue des Martyrs, 38026 Grenoble cedex, France
}
\affiliation{%
\'Ecole Normale Sup\'erieure\\
45 rue d'Ulm, 75005 Paris , France
}

\author{Killian Martineau}%
\affiliation{%
Laboratoire de Physique Subatomique et de Cosmologie, Universit\'e Grenoble-Alpes, CNRS/IN2P3\\
53, avenue des Martyrs, 38026 Grenoble cedex, France
}

\author{Cyril Renevey}%
\affiliation{%
Laboratoire de Physique Subatomique et de Cosmologie, Universit\'e Grenoble-Alpes, CNRS/IN2P3\\
53, avenue des Martyrs, 38026 Grenoble cedex, France
}



\date{\today}
\begin{abstract} 
The idea that, after their evaporation, Planck-mass black holes might tunnel into metastable white holes has recently been intensively studied. Those relics have been considered as a dark matter candidate. We show that the model is severely constrained and underline some possible detection paths. We also investigate, in a more general setting, the way the initial black hole mass spectrum would be distorted by both the bouncing effect and the Hawking evaporation.
\end{abstract}
\maketitle

\section{Introduction}

Although dark matter is a very old problem triggering many studies, no consensual solution has yet emerged. Many experimental searches are being carried out. They are based either on direct (reviews are given in \cite{Censier:2011wd,Gascon:2015caa,Mayet:2016zxu}) or indirect (see \cite{Cirelli:2012tf,Conrad:2014tla,Gaskins:2016cha}) detections.
The production of dark matter particles by accelerators has also been actively considered (see, {\it e.g.}, \cite{Kahlhoefer:2017dnp,Felcini:2018osp} for reviews), without success so far. Quite a lot of ``anomalies" have been registered, for example the overabundance of positrons in cosmic rays \cite{Adriani:2008zr,Accardo:2014lma,Aguilar:2014mma} and the excess of GeV gamma rays from the Galactic Center \cite{TheFermi-LAT:2017vmf}. However none are truly convincing as conventional astrophysical processes can account for the claimed anomalies.\\
Theoretically, many models are being built and there is no point listing them here (one can, {\it e.g.}, see \cite{Plehn:2017fdg} for a brief review). From axions \cite{Klaer:2017ond} to
supersymmetry \cite{Bagnaschi:2015eha} most of them imply new particles.\\

In this article, we focus on the hypothesis proposed in \cite{Bianchi:2018mml}, that is the tunneling of light black holes (BHs) into metastable white hole (WH) relics. In a sense, this would provide a (quantum) gravitational solution to the old dark matter mystery, without relying on modified gravity. We first recall the basics of the model. We then point out several weaknesses ruling out the associated dark matter scenario in conventional cosmological settings. We also underline alternative approaches and investigate the way the initial mass spectrum would be distorted. Finally we consider a possible and potential detection path.

\section{The framework}

\subsection{The bouncing model}

The idea that black holes could tunnel into white holes due to nonperturbative quantum gravitational effects was suggested in \cite{Rovelli:2014cta,Haggard:2014rza,DeLorenzo:2014pta,Haggard:2015iya}. The associated phenomenology was developed in \cite{Barrau:2014hda,Barrau:2014yka,Barrau:2015uca,Barrau:2016fcg,Barrau:2018kyv,Rovelli:2017zoa}. On a dimensional ground, together with general arguments related with the information paradox, the lifetime of a BH with mass $M$ was assumed to be of the order $M^2$. Throughout this work we use Planck units unless otherwise stated.
However, as pointed out in \cite{Christodoulou:2016vny}, this is not supported, at this stage, by actual calculations in full quantum gravity.\\

The initial model was refined in \cite{Bianchi:2018mml}. The idea is the following. The usual semiclassical probability for a black hole to tunnel into a white hole is small for a macroscopic object as it should be of the order of
\begin{equation}
P\sim e^{-M^2}.
\end{equation}
The squared mass term comes from the Euclidean action associated with the considered process. When $M$ approaches unity, the tunneling probability becomes of order 1. In this scenario, before tunneling into a white hole, a black hole first evaporates. From a practical viewpoint this requires its initial mass to be smaller than $\sim 10^{25}$g (approximately the mass of the Moon), otherwise its Hawking temperature is smaller than the one of the surrounding radiation and the evaporation never occurs. When the black hole reaches the Planck mass, it then tunnels into a white hole. The key point is that the formed white hole has a long lifetime and a large interior. Remnants in the form of ``geometric structures" with a small throat and a long tail were, {\it e.g.}, considered in \cite{Banks:1992ba,Giddings:1992kn}.\\

In this model, the usual black hole information paradox is naturally solved as the standard event horizon is replaced by an apparent horizon: information is released after the transition to a white hole. To purify the Hawking evaporation \cite{Page:1993wv}, the remnant has to store information with entropy
\begin{equation}
S\sim M_i^2,
\end{equation}
where $M_i$ is the initial mass of the black hole (before evaporation) and {\it not} the remnant mass \cite{Marolf:2017jkr}. However, due to its small mass, the white hole relic releases information slowly, making it long-lived with a decay time of the order 
\begin{equation}
\tau\sim M_i^4.
\end{equation}
In  \cite{Bianchi:2018mml}, it was shown that an effective metric describing standard black hole radiation followed by a sudden transition to a Planck-mass white hole can be built under natural hypotheses. \\

This scenario is consistent with the fact that old black holes have a large interior volume \cite{Christodoulou:2014yia,Christodoulou:2016tuu}. The geometry outside the black-to-white hole tunneling object is given by a single asymptotically flat spacetime. The Einstein field equations are violated only in two regions: the Planck-curvature one, for which an effective metric that smoothes out the singularity was calculated, and the tunneling one, whose characteristics are known \cite{Christodoulou:2018ryl}. 
A detailed construction defining the black-to-white hole transition amplitude has recently been given in \cite{DAmbrosio:2020mut}.

\subsection{Stable relics}

The Hawking temperature $T_H=1/(8\pi M)$ is extremely small for massive black holes ($T_H/1K=6\times 10^{-8}M_\odot\ /M$) but becomes large for very light ones. The mass loss rate scales as $M^{-2}$, making the whole process explosive. The evaporation mechanism is well described from many different viewpoints and can be considered as consensual (see, {\it e.g.}, \cite{Lambert:2013uaa} for an introduction). There are even indications that it might have been observed in analog systems \cite{Steinhauer:2015saa}. \\

The status of the end point of the evaporation process is, however, much less clear: the standard semiclassical approach breaks down in the Planck region and a nonphysical naked singularity emerges if the usual description is considered without modification. This is why, long before the model considered here emerged, arguments were given favoring of the existence of stable relics (see \cite{Barrow:1992hq,Zeldovich:1983cr,Aharonov:1987tp,Banks:1992ba,Banks:1992is,Bowick:1988xh,Coleman:1991jf,Lee:1991qs,Gibbons:1987ps,Torii:1993vm,Callan:1988hs,Myers:1988ze,Whitt:1988ax,Alexeyev:2002tg} to mention only some references). There are many different ideas in quantum gravity, modified gravity, and string gravity supporting the idea of remnants. A quite generic statement based on known physics was given in  \cite{Giddings:1992hh}: energy conservation, locality, and causality suggest the existence of a huge timescale for the final decay of a black hole.\\

Although no clear consensus exists on the status of BHs at the end of the evaporation process, it is fair to suggest that the existence of relics is somehow expected. A recent review on the pros and cons of stable remnants is given in \cite{Chen:2014jwq}. It is concluded that if relics contain a large interior geometry -- which is the case \cite{Christodoulou:2014yia,Christodoulou:2016tuu} --, they lead to a nice solution to the information paradox and firewall controversy.\\

The possibility that cold dark matter could be made of such relics was first mentioned in \cite{MacGibbon:1987my}. Recently, the possibility that black holes whose remnants which could be dark matter might be formed by the collision of trans-Planckian particles was studied in \cite{Barrau:2019cuo}, following \cite{Saini:2017tsz} and \cite{Conley:2006jg,Nakama:2018lwy}. All this is worth recalling to underline that the long but {\it finite} lifetime of white hole relics considered in this work makes the situation actually more complicated -- to account for dark matter -- than the other scenarios mentioned above where remnants are assumed to be eternal.\\


\section{Production in an inflationary Universe}
 
We first consider the white hole dark matter scenario in the inflationary framework, which we consider to be the current cosmological paradigm. As we shall see, an inflationary stage imposes drastic constraints on the current density of remnants. \\

The most straightforward setting consists of assuming that black holes were formed just at the end inflation, say at (or around) the reheating time \cite{Carr:1975qj}.  For the associated relics to
be still present in the contemporary Universe, their lifetime should to be larger than the Hubble time $t_H$:
\begin{equation}
M_i^4>t_H.
\label{eq2}
\end{equation}
In addition, as they have to be formed by evaporated black holes, one requires
\begin{equation}
M_i^3<t_H,
\label{eq3}
\end{equation}
where $M_i^3$ is the Hawking evaporation time. In a naive deterministic vision, assuming $t_H\sim 13.8$ billion years \cite{Aghanim:2018eyx}, this leads to
\begin{equation}
10^{10}~{\rm g}<M_i<10^{15}~{\rm g}.
\label{eq4}
\end{equation}
It is argued in \cite{Rovelli:2018okm} that this corresponds to typical Hubble masses at reheating, making the scenario convincing. 
The amount of entropy released by the evaporation is, however, severely constrained by nucleosynthesis and this imposes a stringent bound on the number of black holes formed in this mass range. This was studied in details in \cite{Carr_2010,Carr2017,carr2020constraints}. Those constraints must be taken into account so as to check whether they are compatible with the expected density of remnants. As usually done in this framework, we assume that all black holes were formed at the same time and with the same mass $M_i$. Although the resulting constraints could be slightly relaxed by considering a wider mass distribution, the orders of magnitude remain correct. \\

Following \cite{Carr_2010}, let us call $\beta'$ the ratio between the density of black holes and the total density of the Universe at the formation time. Let us call $N_R$ the number of e-folds between the reheating and the contemporary epoch.  The energy density in the form of black holes scales as $a^{-3}$ whereas the energy density of radiation behaves as $a^{-4}$. As the Universe is radiation dominated between the reheating and the equilibrium time, a small initial fraction of black holes is enough to account for current dark matter density.
However, it should be taken into account that black holes contribute only through the mass of the relics -- of the order of the Planck mass -- which makes the situation worse than if they were stable, adding a factor $1/\mathrm{M_i}$. At the reheating, the mass fraction therefore reads:
\begin{equation}
    \beta'= M_ie^{-N_R} \frac{\Omega_{WH}}{\Omega_{r}},
\end{equation}
where the normalized densities of white hole remnants and radiation, $\Omega_{WH}$ and $\Omega_{r}$, are considered today.
As shown in \cite{Carr_2010}, the observational limit is around $\beta' < 10^{-24}$ for $10^{10} \, \mathrm{g}< \mathrm{M_i} < 10^{15} \, \mathrm{g}$. This immediately implies that the current density of white hole remnants must be totally negligible. Assuming $\Omega_{BH}\sim\Omega_{DM}$ would indeed translate in
\begin{equation}
    N_R > 63 + \ln{\left( M_i \right)}.\label{eq:constraint_NR}
\end{equation}
Even with the highest possible reheating temperature, increasing the number of e-folds, this is not compatible with the standard cosmological model. The white hole remnants hypothesis cannot account for any substantial contribution to dark matter in this setting.\\

An alternative cosmological scenario that could be compatible with the formation of a large quantity of black holes is provided by loop quantum gravity (LQG). A pedagogical introductory review can be found in \cite{lqg3}. Although the idea of quasistable white hole relics, studied here, is not rigorously grounded in LQG, there is a natural connection between both frameworks. Indeed, spin foam amplitude for the black-to-white hole transition can, in principle, be explicitly calculated in nonperturbative background-independent quantum gravity \cite{Christodoulou:2018ryl}. When restricted to cosmological symmetries, LQG leads to the so-called LQC paradigm \cite{lqc9,Ashtekar:2015dja}. The most important result is that the big bang singularity is replaced by a regular big bounce \cite{Diener:2014mia}. This opens the exciting possibility that black holes were formed {\it before} the bounce. Although the detailed mechanism is of course highly speculative, it can at least be underlined that the contraction of the Universe makes the growth of inhomogeneities, and therefore the possible BH formation by fluctuations, easier \cite{Barrau:2014kza}. In principle, this helps evading the previous constraint, Eq. \corr{\eqref{eq:constraint_NR}}, as the BHs can now have completed their evaporation before it conflicts with nucleosynthesis.\\ 

The hypothesis that dark matter could be made of relics of black holes formed before the bounce was mentioned in \cite{Rovelli_2018}.  Calling $N_T$ the total number of e-folds between the bounce and the contemporary Universe, the density of relics at the bounce should be $e^{3N_t}$ times higher than in the contemporary Universe. Assuming that the relics contribute substantially to dark matter -- that is $\rho_{WH} \sim 0.3 \rho_{c}$, where $\rho_{c}$ is the critical density -- and choosing a reasonable lower bound $N_T \sim 130$, one immediately concludes that the energy density at the bounce should have been at least $\rho_{WH}e^{390} \approx10^{47}$ in Planck units. This value is not only obviously unphysical, but also inconsistent with the very idea of a relic whose density should be of the Planck scale.\\

This somehow closes the window on the hypothesis of white hole relics within the standard inflationary paradigm.

\section{Production in a matter-bounce Universe}

An interesting alternative to inflation is the matter-bounce scenario which occurs in the matter dominated Universe (see, {\it e.g.} \cite{Lilley_2015} and \cite{br2012matter}). Quite remarkably for the consistency of the model studied here, this might also happen in loop quantum cosmology \cite{WilsonEwing:2012pu}. For a pressureless collapsing universe, the predicted power spectrum of the scalar perturbations after the bounce is scale invariant and the tensor to scalar ratio is negligibly small. A slight red tilt can even be accounted for through an appropriate equation of state.

Let us now assume that black holes were formed at the same mass $\mathrm{M}_i$ and the same time $t_i < 0$ before the bounce (the latter corresponding to $t=0$), and are now present as Planck-mass white hole remnants. Could this model, which is {\it a priori} consistant, be observationally tested? \\

To address this question meaningfully, the lifetime of WHs should be considered stochastic, as a radioactive decay,  rather than deterministic \cite{Barrau:2018kyv}. When we considered the formation of BHs at the reheating, stochastic lifetimes were not relevant. Indeed, this effect would, in principle, allow BHs with a mass below $10^{10}$g to be considered, but the initial density should then be substantially increased and the constraints on their number would prevent WH relics from describing dark matter anyway.

We first consider the hypothesis that a white hole dies by emitting a single quantum at the Planck energy with a period $\tau$. The Universe being essentially transparent at the Planck energy, the flux received by a detector of surface $\sigma$ and of acceptance $\Omega_{acc}$ reads  (neglecting evolution to fix the order of magnitude):
\begin{equation}
   \Phi_{mes} \sim \int_{0}^{R_H} p_{\gamma} \, n_{em} \, \sigma \, \frac{\Omega_{acc}}{4\pi} \, \mathrm{d}r,
\end{equation}
where $n_{em}$ is the emitted flux of the considered source per unit volume and $p_{\gamma}$ is the relative probability of emitting a photon. Since the process can be assumed democratic, as in the case of the Hawking evaporation, $p_{\gamma}$ is simply given by $2/n_{SM}$, where $n_{SM}$ is the number of degrees of freedom of the standard model. Obviously, one could imagine that new degrees of freedom exist at higher energies, but this does not dramatically change the picture. 

Calling $ \mathrm{d}N_{ev} $ the number of events in a volume d$ V $ during a time d$ t $, one has:
\begin{equation}
    \mathrm{d}N_{ev}= p_{\gamma}\,\left(\rho_{N}\,\mathrm{d}V\right)\,\left(\frac{1}{\tau}\,\exp{(-(t-t_{i})/\tau)}\,\mathrm{d}t\right),
\end{equation}
where $t_i$ is the formation time and $\rho_{N}$ is the number density of relics. Gathering the different elements of the straightforward calculation, now along a null geodesics, one is led to
\begin{equation}
     \Phi_{mes}= \Phi_{0}\, \int_{0}^{1} \frac{\mathrm{d}u}{a(u)} \exp{\left(-\frac{t_{H}}{\tau}\, u\right)}
\end{equation}
with $u=t/t_{H}$, $t_H$ the Hubble time and  $\Phi_{0} $  a number which does depend on the parameters $ t_{i} $ and $ M_{i} $.\\

The situation can be summarized as follows. In the scenario considered here, black holes have evaporated and turned into white hole remnants before the bounce. However, in itself, this does not put {\it any} restriction on the initial mass $M_i$ and formation time $t_i<0$ (other than the trivial requirement that the absolute value of formation time is larger than Hawking time for the corresponding mass, $M_i<(-t_i)^{1/3}$). By construction, the current density of white hole relics is indeed assumed to be the one of dark matter. As the decay process is stochastic (the remaining density of relics never reaches exactly zero), it is always possible to normalize the initial density so as to fulfill the requirement $\Omega_{WH}=\Omega_{DM}$, regardless of the initial mass and formation time. This makes the hypothesis naturally self-consistent, except in the specific case where the initial mass is very small and the formation time very close to the bounce, reviving the problem of a trans-Planckian density at the bounce. \\

An interesting point consists of comparing the expected flux with the available measurements. The numerical prefactor $\Phi_{0}$ reads
\begin{equation}
\Phi_{0}\sim 2 \cdot 10^{15}\,\frac{1}{\tau}\,\exp{\left(\frac{t_{i}}{\tau}\right)}.
\end{equation}
The Auger observatory \cite{Aab:2016vlz}, with a very large active surface of the order of 3000 km$^2$, has not yet detected any ``Planck-energy" event. When gathering all the numbers and performing the integration, this can be translated into a bound
\begin{equation}
    M_i  >  10^{12} \, \mathrm{g},
\end{equation}
which reduces the initial parameter space.
It is worth wondering if this limit could be improved in the future. The most extreme idea to search for cosmic rays with considerable energies is probably the use of Jupiter as a huge calorimeter to detect particle showers \cite{Rimmer:2014cia}. The effective surface becomes 4 orders of magnitude higher that the one of the Auger observatory. As the mass appears in the flux with a power 1/4 this would only rise the limit to $M_i  >  10^{13} \, \mathrm{g}$.\\

Obviously, the way in which a metastable WH remnant decays is unknown. We shall now consider a toy model as different as possible from the one previously studied. Instead of assuming that the white hole disappears by the emission of a single Planck-energy quantum, we consider a continuous emission of low-energy radiation during its entire lifetime. It is meaningful to consider an $E^{-2}$ differential spectrum. First, because it is quite usual and nearly generic in astrophysical processes (although it is fair to underline that no ``Fermi-like" process is expected in this model). Second, because it has the strong heuristic advantage of being normalized by the logarithm of the ratio of the maximum to minimum energies, $E_1$ and $E_2$, thus avoiding the requirement to fix them precisely. 

Requesting that the WH relics are the main component of dark matter leads to an emission density at energy $E$: 
\begin{equation}
    n_{em}= \rho_{DM} \frac{1}{\tau_{WH} \, \ln{(\mathrm{E}_2/\mathrm{E}_1})} \frac{1}{\mathrm{E}^2}.
\end{equation} 

In this case,  the relics emit a continuous signal during their entire lifetime until the full disappearance. By construction, the model is here somehow deterministic and the requirement that the white hole relics  explain the dark matter reads  $t_H-t_i < \tau+\tau_{BH} \simeq \tau$  where $\tau_{BH}$ is the Hawking time. In the previous case, the stochastic nature of the emission makes it always possible to find a normalization such that the current density approaches the dark matter density. This is not true here and this is the reason why both $M_i$ and $t_i$ are now relevant.

The idea, in this case, consists of comparing the flux emitted by the population of relics with the diffuse astrophysical background.  This translates into a lower limit on $\tau$ -- otherwise too much radiation would be emitted -- which itself leads to $M_i> 10^{11}\, \mathrm{g}$, a result remarkably similar to the constraint obtained in the previous and radically different scenario. The additional constraint on the formation time reads $ | t_i | <\tau_{WH} -t_H$ that is, to a good approximation, $ | t_i | < M_i^4$. This is of the order of $10^{22} \, \mathrm{s} $ for the lowest authorized $M_i$ value, which is 5 orders of magnitude larger than the Hubble time. It means that the time interval allowed for black holes to be produced in the contracting branch of the Universe, possibly contributing now to dark matter through white hole relics, is huge (although finite) for any allowed initial mass.

\section{A few remarks on the mass spectrum of primordial black holes}

We now switch to the bouncing model, mentioned in the Introduction, where the bouncing time is $\tau=kM^2$  \cite{Rovelli:2014cta,Haggard:2014rza,DeLorenzo:2014pta,Haggard:2015iya} with the aim of clarifying how it affects the BH mass spectrum.\\

The shape of the initial mass spectrum of primordial black holes highly depends on the production mechanism and is therefore not known. When assuming that BHs are directly formed by density fluctuations, the initial mass spectrum is given by
\begin{equation}
\frac{{\rm d}n}{{\rm d}M_i}=\alpha M_i^{-1-\frac{1+3w}{1+w}},
\label{spec}
\end{equation}
where $w=p/\rho$ is the equation of state of the Universe at the formation epoch \cite{Carr:1975qj}. Focusing on other formation scenarios, in particular, phase transitions \cite{Jedamzik:1999am,Rubin:2000dq}, can lead to very different spectra. The question we want to address here is the one of the distortion of the mass spectrum induced by both the effect of the Hawking evaporation and that of the hypothetical bounce.\\

It is well known and easy to show that the evaporation will make the contemporary spectrum roughly proportional to $M^2$ up to $M_*$ \cite{Barrau:2002mc}, the mass corresponding to an evaporation time equal to the time difference $\Delta t$ between the formation and detection times (that is the Hubble time if one considers BHs formed in the primordial Universe). Let us go beyond this crude estimate and clarify the situation. 

Let $ f_i(M) $ be the initial spectrum, just at the end of inflation in typical models, such that dn = $ f_i(M) $ dM. The only required hypothesis is that $ f_i(M) $ is smooth. Black holes with initial masses below $M_*$ do not contribute to the contemporary spectrum. On the other hand, black holes with masses well above $M_*$ have not yet significantly evaporated and their mass spectrum is the same as the initial one. The full relevant part of the contemporary spectrum, that is, $M\leq M_*$, is determined by the behavior of black holes with initial masses slightly above $M_*$. The mass loss rate is obtained by integrating $E\phi(E)$, where $\phi(E)$ is the instantaneous differential Hawking spectrum,
\begin{equation}
\phi(E)
=\Gamma_s\left[\text{exp}\left(8\pi M E\right)-(-1)^{2s}\right]^{-1},
\label{hawk}
\end{equation}
per state of angular momentum and spin $s$. The absorption coefficient $\Gamma_s$ is the probability that the particle would be absorbed if it were incident in this state on the black hole and $E$ is the energy of the emitted quantum.  This leads to $dM/dt\propto -M^{-2}$ and, therefore, a BH of initial mass $M_{i}$ reaches, after a time t, a mass $M_{f}= M_i {\left( 1 - (M_{\ast}/M_{i})^3 t/\Delta t \right) }^{1/3} $. Let us consider $M_i = M_{\ast} (1+ \epsilon)$ so that $M_f = M_{\ast} (3\epsilon)^{1/3}$. A straightforward calculation shows that, denoting $f(M)=f_i(M_i) \frac{\mathrm{d}M_i}{\mathrm{d}M} $, the contemporary spectrum,
\begin{eqnarray}
    f(M)     &=& \left( f_i(M_{\ast}) + M_{\ast}\epsilon \frac{\mathrm{d}f_i}{\mathrm{d}M_i}(M_{\ast})+ O(\epsilon^2) \right)\nonumber \\ 
    &\times& \left({\left( \frac{M}{M_{\ast}} \right)}^2 + O(M^6) \right),
\end{eqnarray}    
which gives, after a first-order expansion in $ \epsilon $, and using  $3\epsilon=(M/M_{\ast})^3$,
\begin{equation}
   f(M) = f_i(M_{\ast}) \, {\left( \frac{M}{M_{\ast}} \right)}^2 + \frac{1}{3} M_{\ast} \frac{\mathrm{d}f_i}{\mathrm{d}M_i}(M_{\ast}) {\left( \frac{M}{M_{\ast}} \right)}^5 + O(M^6).
\end{equation}
This shows that the low-mass spectrum is indeed proportional to $M^2$ and this clarifies the higher-order corrections.\\

Let us now add the bouncing effect. It should be treated stochastically otherwise the result is trivial : the spectrum would be unchanged for $M>M_B$, where $M_B$ is the mass corresponding to a bouncing time equal to $\Delta t$, and vanishing for $M<M_B$. Focusing on the relevant dynamics, the result is qualitatively as follows. Black holes with current masses below $M_*$ all had nearly the same initial mass of the order of $M_*$. Their number density due to the bouncing effets is therefore reduced by the {\it same} factor $e^{-\Delta t/\tau(M_*)}$. This translates to a simple damping of the spectrum amplitude. The situation is different above $M_*$. In this case, the damping factor depends on the mass and the spectrum is multiplied by a factor $e^{-\Delta t/\tau(M)}=e^{-\Delta t/(kM^2)}$. \\

To summarize, the resulting contemporary spectrum, taking into account both the Hawking evaporation and the stochastic bouncing effect, is proportional to $M^2$ when $M<M_*$, but instead of reaching (nearly) the initial mass spectrum at $M_*$, it is now damped by a factor $e^{-\Delta t/(kM_*^2)}$. When $M>M_*$ the correction to the initial spectrum becomes mass dependent and converges to unity for $M\rightarrow \infty$.\\

Let us be a little more specific on the details by focusing on the interesting region with BHs of masses close to $M_*$. Defining $\beta = \Delta t/k \, {(1/M_{\ast})}^2$, such that $ \alpha \approx \beta \, (1-2\epsilon+O(\epsilon^2)) $, where $\alpha = \Delta t/\tau$, it can be shown that 
\begin{eqnarray}
    f(M) &=& e^{-\alpha}\, f_i(M_{i}) \, {\left( \frac{M}{M_{\ast}} \right)}^2 = e^{-\beta}\,e^{2\beta\,(\epsilon+O(\epsilon^2))}\, \nonumber \\
    &\times& \left(f_i(M_{\ast}) + M_{\ast}\epsilon \frac{\mathrm{d}f_i}{\mathrm{d}M_i}(M_{\ast}) + O\left(\epsilon^2\right) \right) \, \frac{M^2}{M_{\ast}^2},
\end{eqnarray}
which leads, after a first-order expansion in $ \epsilon $, and using that $(3\epsilon)=(M/M_{\ast})^3$, to 
\begin{widetext}
\begin{equation}
f(M)= e^{-\beta} \left( f_i(M_{\ast}) \, {\left( \frac{M}{M_{\ast}} \right)}^2+ \left(\frac{2}{3}\beta \, f_i(M_{i}) + \frac{1}{3} M_{\ast} \frac{\mathrm{d}f_i}{\mathrm{d}M_i}(M_{\ast}) \right) {\left( \frac{M}{M_{\ast}} \right)}^5 \right) + O(M^6)
\end{equation}
\end{widetext}

This establishes the shape of the mass spectrum when taking into account both the "perturbative" Hawking evaporation -- seen here as a kind of dissipative process -- and the ``nonperturbative" quantum bounce. 

\section{Binary systems}

It is finally worth considering the interaction of relics. Whatever the detailed cosmological setting considered, compatible with the idea that WH remnants constitute most of the dark matter, binary systems could form. Although nothing can enter a WH, its neat ``gravitational force" exerted on a distant object remains {\it attractive}. In principle, forming binary systems of white holes is possible \cite{Barrau:2019cuo}. It should, however, be kept in mind that the current number density of Planck-mass relics necessary to describe dark matter is extraordinary small, of the order of one per billion cubic kilometers. \\

Following \cite{Sasaki:2016jop} and \cite{Nakamura:1997sm}, we estimate the probability of coalescence in the interval $(t,t+dt)$ to be
$$
dP=
    \frac{3}{58} \bigg[ -{\left( \frac{t}{ \frac{3}{170}\alpha^4} \right)}^{3/8}
+{\left( \frac{t}{T \frac{3}{170}\alpha^4} \right)}^{3/37} \bigg] \frac{dt}{t},
$$
with  
$$
\alpha={\left( \frac{1}{\rho_{rel}(z_{\rm eq})} \right)}^{1/3},
$$
assuming that the mass of the relic is equal to the Planck mass and calling $\rho_{rel}(z_{\rm eq})$ the relative density of WH remnants at the equilibrium time.
Under the assumption that most relics were already formed at this epoch, the coalescence rate is of the order of $10^{-45}{\rm m}^{-3}{\rm s}^{-1}$. 

This should of course be taken with extreme care. To the best of our knowledge, the details of the merging of two WHs is not known. It is even probable that, because of the structure of WHs, a merging in the usual sense does {\it not} happen. However, it is clear that binary systems emitting gravitational waves are possible and even expected when WHs are in the appropriate kinematical conditions. \\

Those gravitational waves are undetectable, both in amplitude and in frequency. But the object resulting from the ``coalescence" might be detectable if it were to emit a  Planck-energy quantum. This makes sense as it is naively expected to relax to the state of metastable Planck-mass WH relic. 
If considering the Auger detector \cite{Covault:2019mdm}, this flux would be too small for a detection. For a detector of the Euso \cite{Inoue:2009zz} class, the rate reaches an event per decade. When considering long run projects planning to use Jupiter as a cosmic-ray detector \cite{Rimmer:2014cia}, as mentioned in the previous sections, the expected flux becomes of the order of a dozen events per year, which is experimentally meaningful.

\section{conclusion}

The idea that white hole relics could be the main component of dark matter is an appealing one, which is grounded in quite convincing quantum gravity inspired arguments.  Unquestionably, the fact that black holes might turn white, either before they undergo most of the Hawking evaporation -- with a bouncing time of the order $M_i^2$ -- or at the end of the evaporation -- with a surviving time $M_i^4$ -- is worth considering and could be a major prediction for models beyond general relativity. A possible connection with dark matter is a natural hypothesis. However, we have shown that the model is severely constrained if $\Omega_{WH}$ is to be of the order of $\Omega_{DM}$:

\begin{itemize}
\item The production of enough black holes at the end of inflation is ruled out, mostly to avoid too much entropy  being released and conflicting with nucleosynthesis.
\item The production of enough black holes before inflation is ruled out due to the unphysical associated energy density at the formation time.
\item The production of black holes in the contracting branch of the Universe in bouncing cosmological models without inflation -- which are already challenging to build -- is possible, but in a restricted parameter space. Considering two very different scenarios for the details of the decaying process, we have shown that the initial mass should  be higher anyway that the naive lower bound.
\end{itemize}

We have also clarified the shape of the mass spectrum of black holes when taking into account both the Hawking evaporation and the bouncing hypothesis. Finally, a long run prospect for detecting WH dark matter has been suggested. 

\bibliography{refs}

 \end{document}